\documentclass[manuscript]{acmart}
\AtBeginDocument{%
  \providecommand\BibTeX{{%
    \normalfont B\kern-0.5em{\scshape i\kern-0.25em b}\kern-0.8em\TeX}}}

\copyrightyear{2023} 
\acmYear{2023} 
\setcopyright{acmlicensed}\acmConference[CHI '23]{Proceedings of the 2023 CHI Conference on Human Factors in Computing Systems}{April 23--28, 2023}{Hamburg, Germany}
\acmBooktitle{Proceedings of the 2023 CHI Conference on Human Factors in Computing Systems (CHI '23), April 23--28, 2023, Hamburg, Germany}
\acmPrice{15.00}
\acmDOI{10.1145/3544548.3581309}
\acmISBN{978-1-4503-9421-5/23/04}

\newcommand{\sys}[0]{{{\it CoPracTter}}}

\def\markup{0}

\if\markup1
\newcommand{\rv}[1]{{\leavevmode\color{blue}#1}}
\else
\newcommand{\rv}[1]{#1}
\newcommand{\st}[1]{}
\newcommand{\sout}[1]{}
\fi

\usepackage{float}

\begin{document}

\title[CoPracTter: An Online Support Tool for Coping with Stuttering]{CoPracTter: Toward Integrating Personalized Practice Scenarios, Timely Feedback and Social Support into An Online Support Tool for Coping with Stuttering in China}


\author{Li Feng}
\affiliation{
  \institution{Computational Media and Arts Thrust}
  \institution{The Hong Kong University of Science and Technology (Guangzhou)}
  \city{Guangzhou}
  \country{China}
}
\email{lfeng256@connect.hkust-gz.edu.cn}

\author{Zeyu Xiong}
\affiliation{
  \institution{Computational Media and Arts Thrust}
  \institution{The Hong Kong University of Science and Technology (Guangzhou)}
  \city{Guangzhou}
  \country{China}
}
\email{zxiong666@connect.hkust-gz.edu.cn} 

\author{Xinyi Li}
\affiliation{
  \institution{School of Engineering}
  \institution{The Hong Kong University of Science and Technology}
  \country{Hong Kong SAR, China}
}
\email{xliij@connect.ust.hk} 

\author{Mingming Fan}
\orcid{0000-0002-0356-4712}
\affiliation{%
  \institution{Computational Media and Arts Thrust}
  \institution{The Hong Kong University of Science and Technology (Guangzhou)}
  \city{Guangzhou}
  \country{China}
}
\affiliation{%
  \institution{Division of Integrative Systems \& Department of Computer Science and Engineering}
  \institution{The Hong Kong University of Science and Technology}
  \country{Hong Kong SAR, China}
}
\authornote{Corresponding Author}
\email{mingmingfan@ust.hk}
\renewcommand{\shortauthors}{Li Feng et al.}

\begin{abstract}
Stuttering is a speech disorder influencing over 70 million people worldwide, including 13 million in China. It causes low self-esteem among other detrimental effects on people who stutter (PwS). Although prior work has explored approaches to assist PwS, they primarily focused on western contexts. In our formative study, we found unique practices and challenges among Chinese PwS. We then iteratively designed an online tool, CoPracTter, to support Chinese PwS practicing speaking fluency with 1) targeted stress-inducing practice scenarios, 2) real-time speech indicators, and 3) personalized timely feedback from the community. We further conducted a seven-day deployment study (N=11) to understand how participants utilized these key features. \rv{To our knowledge, it is the first time such a prototype was designed and tested for a long time with multiple PwS participants online simultaneously.} Results indicate that personalized practice with targeted scenarios and timely feedback from a supportive community assisted PwS in speaking fluently, staying positive, and facing similar real-life circumstances. \st{Finally, we present design implications for better supporting PwS.} 
\end{abstract}




% A "teaser" image appears between the author and affiliation
% information and the body of the document, and typically spans the
% page.
%   \includegraphics[width=\textwidth]{figure/overview.png}
%   \caption{Three main components in \sys{}: (A) Examples of practice scenario videos which were used in the user study; (B) Real-time Feedback in the practice phase: facial expression, audio transcription, speech rate, pitch and volume; (C) Social support including encouraging words and timely personalized feedback from other PwS in the community.}
%   \Description{The overview of main components in the prototype: Left: three example video screenshots of simulated practice scenarios used in the user study; Middle: the display of real-time feedback: facial expression, audio transcription, speech rate, pitch and volume; Right: a paragraph of encouraging words and comments from other PwS.}
%   \label{fig:overview}

\maketitle

\section{Introduction}
% definition and negative impacts of Stutter
Stuttering, also known as stammering, is a speech disfluency that impacts around 1\% of the global population \cite{facts2022}. The cause of the disorder varies across people, but it is identified by recurrent interruptions of the natural flow of speech or part-word repeats \cite{Maguire1997pet}. Although many people do not perceive stuttering to be a disability, stuttering could have a variety of negative repercussions on the everyday lives of people who stutter (PwS). For example, stuttering may induce anxiety or shame in both children \cite{smith2017anxiety} and adults \cite{craig2009impact}, which may be linked to the fact that PwS generally have lower levels of self-perceived communication competence than those who do not stutter \cite{werle2021preliminary}. In addition, the general public may have less trust in PwS and perceive them as being less competent, mature or knowledgeable, which could lead to discrimination and social devaluation in the workplace \cite{alqhazo2017discrimination, plexico2019influence}. Some PwS who have encountered stigma may feel they will continue to face stigma in the future, which might have a detrimental effect on their mental health and self-esteem \cite{boyle2018enacted}. As a result, it might create long-term difficulties for PwS in avoiding potentially embarrassing circumstances, such as public speaking, forming lasting relationships, and finding a job.

Within the HCI and accessibility communities, researchers have investigated different ways to assist PwS, including exploring design considerations for inclusive speech interfaces for PwS, helping PwS write scripts that they could speak more fluently, diagnosing the degree of stuttering, recording stuttering situations for SLP to diagnose, and uncovering the type of support that PwS would want to have. \rv{While informative, these approaches were limited in two ways. First, they were mostly designed and evaluated with PwS living in western cultures. Cultures could influence stuttering behaviors \cite{leith1975cultural}. For instance, black stutterers are more likely to have secondary characteristics (e.g. speech modifiers) that could covert prolongations and repetitions than white stutterers. Moreover, people's attitudes toward stuttering tend to vary across cultural contexts. Prior work reported significant differences in people's attitudes toward stuttering among British, Arab, and Chinese, such as perceived reasons for stuttering, assistance and empathy for PwS \cite{ustun2021cultural}. And people's attitudes toward stuttering could affect the experiences, personal identity and mental health of PwS \cite{McNaney2018stammerApp, daniels2006qualitative, blood2003preliminary}. }However, little is known about the experiences of PwS in China, which has 13 million PwS. 
The second limitation lies in the type of support. Prior work mostly focused on offering individual speaking practicing support with minimum peer support \cite{McNaney2018stammerApp}. For example, StammerApp merely provided links to websites \cite{McNaney2018stammerApp}, which required PwS to read through texts on websites that did not offer feedback on their own practice sessions. Informed by the two limitations, we sought to take a step further to tackle these limitations by first answering a research question(RQ1): \textbf{\st{How do Chinese PwS stutter in daily lives?} \rv{What are the stuttering situations that PwS in China encounter? What are their current workarounds and challenges?}}

% Weidner et al. \cite{weidner2017comparison} found the stuttering stereotype is shaped by the culture in which the individual has been reared.

% online observation and interview findings
\rv{To answer RQ1, we conducted a formative study in which we first analyzed the content posted in four representative and reliable PwS online communities frequently visited by PwS in China (Zhihu\footnote{Zhihu: https://www.zhihu.com/, has 30 million daily active users}, Weibo, Baidu Tieba and Douban). Then we conducted semi-structured online interviews with 12 PwS from different regions of China. From the formative study, we identified the need for personalized speech practice and timely feedback on their practice from their peers due to the shortage of SLP and assistive mobile apps for PwS to practice speaking. Informed by the stuttering situations and the need for personalized timely feedback from their peers, we derived five design considerations (DCs)}

Based on the DCs, we further designed and implemented a mobile assistive app, \sys{}. \sys{} offers stutter-triggering scenarios for PwS to practice, with personalized timely feedback. We then used it to investigate the effectiveness and user experience of these features by answering the second research question (RQ2): \textbf{How are personalized practice scenarios, timely targeted feedback, and social support used by PwS in their daily speaking practice?}

To answer RQ2, We conducted a 7-day deployment study where 11 PwS participated online simultaneously (N=11). \rv{To our knowledge, this is the first deployment study involving multiple PwS participants simultaneously. Participants were required to use it for speaking practice and giving others timely feedback every day. By the end of the study, they were interviewed about their experiences of using the prototype.} The results indicated that personalized stutter-triggering video simulation tasks with timely feedback could aid PwS to practice speaking fluency with appropriate pressure. All PwS participants found subjective comments from other PwS to be the most beneficial while objective real-time speech-related feedback is also helpful (e.g. speech rate, facial expression et al.). Participants have different preferences for the feedback types and suggested potential ways to improve the design of some features (e.g. transcription, speech rate et al.). 

\sout{To answer RQ1, we first analyzed the content posted in major PwS online communities in China and then conducted interviews with Chinese PwS (N=12) to investigate the strategies, challenges and user needs of PwS community in China. We identified the following key insights into Chinese PwS situations: (1) Many PwS have negative experiences due to their stuttering and want to eliminate them; (2) Stuttering situations are highly personalized and individualized targeted practices are more appreciated than general practices; (3) One key factor for improving the stuttering situation is to continue speaking daily and gaining timely and targeted feedback; (4) The majority of PwS anticipate a more encouraging online community to support and assist each other.
To help PwS maintain their self-esteem and mental wellness, encouraging words and a supportive community could help \cite{trichton2018peer, raj2017psychosocial}. To efficiently practice speaking, targeted practice, which is more personalized and focuses on situations in that they tend to stutter, should be leveraged instead of general practice \cite{ericsson1993role, ericsson2016peak}. Besides keeping practicing every day, it is also important to receive timely targeted feedback and advice from a friendly PwS community and quantitative evaluation from speech-related indicators. Therefore, we are motivated to investigate our second research question (RQ2): \textbf{How can personalized practice scenarios, timely targeted feedback and community support benefit PwS for their daily speaking practice?}
}

In summary, we made the following contributions:
\begin{itemize}
    \item We uncovered common stuttering scenarios, workarounds and challenges of PwS in China through a formative study including online PwS community analysis and interviews. 
    \item Informed by the formative study and through an iterative design process with PwS, we designed and implemented an interactive prototype, CoPracTter, that allows PwS to practice speaking in practical scenarios to induce varying stress levels, review AI-extracted speech features while practicing, share practice recordings with the community, and receive timely and personalized feedback from the community.
    \item We conducted a 7-day deployment study with a community of PwS simultaneously online to understand how they used and perceived the effectiveness of these key features and presented design implications. \rv{According to our knowledge, it is the first time such a prototype has been designed and evaluated in a long-term, multi-user simultaneous online deployment study.} 
\end{itemize}

\section{Related Work}
\rv{\subsection{PwS Suffer from Anxiety and discrimination}}
Stuttering could have many negative impacts on PwS. Prior work revealed that stuttering could rise the anxiety level of individuals \cite{iverach2014social}. PwS were found to have higher anxiety levels than people who do not stutter (PwNS) at all ages: children \cite{smith2017anxiety}, adolescents  \cite{mulcahy2008social} and adults \cite{craig2009impact}. A community cohort study with 843 eleven-year-old children found that the group with persistent stuttering had considerably higher anxiety than the group with recovered stuttering and non-stutter controls \cite{smith2017anxiety}. A comparative study reported that stuttering adolescents have significantly higher anxiety than non-stuttering controls \cite{mulcahy2008social}. For adults who stutter (AwS), stuttering could also have negative impacts on their social interaction abilities \cite{craig2009impact}, which may be linked to the fact that PwS generally have lower levels of self-perceived communication competence than those who do not stutter \cite{werle2021preliminary}. One of the potential influencing factors for their anxiety and social interaction ability may be the public's attitude. An online survey about internalized feelings and discrimination experienced by PwS revealed that most PwS have been treated unfairly by the public, which made them feel annoyed and embarrassed \cite{alqhazo2017discrimination}. Therefore, PwS, especially young adolescents, would try to hide their stuttering \cite{blood2003preliminary}. An interview study revealed that the quality of life for PwS was significantly reduced by stuttering in four main aspects: vitality, social function, emotional function, and mental health \cite{craig2009impact}. More detrimentally, many prior works found that the impact of stuttering on PwS often lasts for a long term. Boyle et al. conducted an online survey to investigate the enacted stigma, felt stigma, and global mental health of PwS \cite{boyle2018enacted}. They found that many PwS who have experienced stigma tend to anticipate that they will continue to suffer stigma for the long term in the future. In addition to social lives, stuttering also impacts PwS greatly in the workplace. Plexico et al. conducted an online study with both PwS and people who do not stutter \cite{plexico2019influence}. They found that the participants who stutter differ from people who do not stutter in terms of job satisfaction, discrimination, and vigilance.

\sout{Stuttering (or stammering) is caused by various reasons (psychological~\cite{BOYLE2009201, boyle2013psychological}, genetic~\cite{drayna2011genetic, yairi1996genetics, kraft2012genetic}, etc.) and has numerous personalized behaviors. As Lea et al. \cite{lea2021sep} mentioned in their work, there are mainly five categories of stuttering: prolongations (a syllable is prolonged for several seconds e.g. m----mommy), sound repetitions (repeated syllables e.g. pr-pr-prepared), word/phrase repetitions (e.g. I made, made dinner), injections (filler words such as ‘um’ or ‘uh’), and block (gasps for air or stuttered pauses).}

\rv{\subsection{Experience of PwS in China}
Several studies have reported that compared to western societies, PwS in China are more likely to have negative experiences and attitudes toward their stuttering. A social study revealed that Chinese people tend to have more negative ideas about stuttering than people in western nations, and Chinese PwS lacked the typical optimism for pursuing any career they desired~\cite{ip2012stuttering}. An online survey study revealed significant variations between British, Arab, and Chinese attitudes toward stuttering, including their attributions of stuttering etiology, their role in assisting PwS, and their empathy for PwS \cite{ustun2021cultural}. Particularly, Chinese people have more negative attitudes towards these issues, which may result in Chinese PwS tending to hide their stutter and not seeking help from others. In addition, the listener's behavior when listening to PwS speaking English is also different across cultures, which may result in different workarounds for PwS in different cultures. Zhang et al. provided empirical evidence that Chinese listeners spent more time on the speaker's background and less on the eyes and nose than Americans \cite{zhang2012culture}. Jin et al. pointed out that little is known about speech-language pathology in China, and scientific research on the behavior of stutterers based on this science are still in its infancy~\cite{ming2001public}. This aligns with the findings of a survey study that people in Rio de Janeiro and Belgium all treated stuttering more seriously and devoted more resources to speech-language therapy \cite{de2008public}. Although prior works have revealed many differences in stuttering between western societies and China and the lack of stuttering-related research in China, none of them investigated the Chinese PwS user experiences and needs in detail. Therefore, we conducted a formative study to investigate potential differences in stuttering situations and challenges for PwS in China
}

\subsection{Supporting Tools for PwS}
Within the HCI and accessibility community, researchers have been devoted to investigating ways to assist PwS. \rv{Vigot et al. investigated the effectiveness of different types of auditory feedback in reducing stuttering \cite{Voigt2014poster}. Specifically, they deployed delayed auditory feedback on smartphones and found its effectiveness limited due to the native latency of the smartphones. A speech script writing assistance tool was also developed, which could intelligently replace stutter-trigger words with easy-to-pronounce ones with similar meanings \cite{ghai2021fluent}. Much work in the community focused on facilitating SLP to identify their patients' stuttering context remotely in time as well as aiding PwS to recognize their own progress. Chandra et al. investigated the usage of social robots in the stuttering clinic \cite{chandra2022opportunities}. They presented eight scenarios that can be adapted for stuttering intervention with social robots, such as cooperation between social robots and a single user, music modeling and cooperative games. The iAmS \cite{madeira2013building} is a prototype allowing PwS to register their stutter-related situations in a regulated workflow. The system will notify the SLP associated with the specific user about the detailed information of his stuttering situation in time so that they could identify the context better. Later, the prototype was iterated with personalization features and renamed as BroiStu \cite{demarin2015impact}, which was evaluated with a user study involving both SLP and PWS to rate each feature. The design proved to be helpful for both parties. The implementation of the personalization aspect was presented and specifically evaluated by Madeira et al. in their later work \cite{madeira2018personalising}. The effects of these self-reflective applications are highly dependent on SLPs. While the therapy from SLP is expensive \cite{meyer2011speech}, in China, there is a dearth of SLPs with expertise in stuttering, which results in insufficient SLP therapy and a contradiction between PwS and SLP \cite{slpWeb}. Therefore, it is crucial to develop a strategy to assist PwS in improving their speech fluency more independently without relying on SLPs. 
}

\sout{Many prior works have designed self-reflective mobile applications to assist PwS to record their stuttering situations in time and maybe report to their Speech-Language Therapist (SLT) or Speech-Language Pathologists (SLP). The iAmS \cite{madeira2013building} is a software that allows users to register their stutter-related situations and generates a report that illustrates the evolution of the user's stuttering conditions. In addition to promoting daily self-monitoring of speech by users, it enables speech therapists to monitor and receive reports on the speech of their patients. BroiStu \cite{demarin2015impact} is an iterated version of iAmS, which includes four new modules in design: personalization, exercises, voice interaction and communications. However, the new features were not deployed. Demarin et al. evaluated the design with a user study involving both SLP and PWS, where they were asked to rate each feature. The design proved to be helpful for both parties. Madeira et al. \cite{madeira2018personalising} then presented the implementation of personalization aspects in BroiStu and evaluated it with a preliminary experiment. However, the effects of these self-reflective applications are still dependent on SLTs. In China, there is a dearth of SLPs with expertise in stuttering, and the therapy is rather expensive for many PwS \cite{meyer2011speech}. Therefore, it is crucial to develop a strategy to assist PwS in improving their speech fluency without the assistance of SLPs.}

\sout{The importance of Understanding user experiences of PwS is also emphasised as one key challenge by Clark et al. \cite{Clark2020Speech}.They stated that adopting methods like participatory design and co-design including people with diverse speech patterns would be beneficial for the research process. McNaney et al. presented the user-centered design, development and evaluation of a PwS supporting mobile application - StammerApp \cite{McNaney2018stammerApp}. By engaging PwS in the design process, they identified the challenges and barriers that PwS experience day-to-day and provided design recommendations for tools to support PwS in everyday interactions. However, their design mainly focused on the self training and reflection process, and they did not investigate the effect of the personalized feedback from other PwS in a supportive community where everyone is willing to practice and communicate with others to improve theirs speech fluency.}

\rv{
When designing tools to support PwS to improve their speech fluency more independently, it is important to involve PwS in the design process to better cater to their needs \cite{Clark2020Speech}. McNaney et al. conducted a survey study with PwS to understand the barriers PwS face in their daily lives and their need for digital support \cite{McNaney2018stammerApp}. They also further developed a PwS-supporting mobile application - StammerApp, where PwS can find SLP, practice speaking, take journals and connect with other PwS. However, their design mainly focused on supporting self-training using six scenarios without personalized real-time feedback and advice from others, which would constrain the training effect. Meanwhile, for the aspect of linkage with other PwS, they only provided links to the generic support groups and forums, yet no personalized real-time feedback or advice from other PwS was provided. Our prototype, \sys{}, integrated different types of personalized feedback with various stutter-trigger situations for PwS to practice within an online community. 
}
\section{Formative Study}
To answer RQ1, we conducted a formative study including two parts: (1) an analysis of posts in popular Chinese PwS online forums to obtain a preliminary understanding of the stuttering experiences and user needs of PwS and (2) a semi-structured interview (N=12) to investigate their needs in detail. The findings from the formative study highlight the user needs to guide the design of a prototype which can help PwS practice speech fluency in personalized scenarios and obtain timely feedback within a supportive community. Based on these findings, we then performed an iterative design process with 12 PwS to better address their needs.

\begin{table*}[ht]
\scalebox{0.9}{
 \begin{tabular}{ |c c c c c| } 
 \hline
 \textbf{Id} & \textbf{Gender} & \textbf{Age} & \rv{\textbf{Location}} & \textbf{Self-description of Stuttering Situations} \\ 
 \hline
     1 & male & 22 & \rv{Nanchang} & I sometimes repeat the first word and could not move on.\\ 
     2 & male & 17 & \rv{Chengdu} & Sometimes I want to speak but I am afraid to, especially when making friends.\\ 
     3 & male & 23 & \rv{Guangzhou} & I have difficulty pronouncing some syllables. I will stop in the middle of a sentence.\\ 
     4 & male & 23 & \rv{Nanchang} & I have difficulty pronouncing some syllables, especially the first syllables in a sentence. \\ 
     5 & male & 23 & \rv{Hangzhou} & I repeat some words when I say something important.\\ 
     6 & female & 19 & \rv{Yichun} & I have difficulty pronouncing some syllables.\\ 
     7 & male & 35 & \rv{Fangchenggang} & I could not open my mouth when I stutter.\\ 
     8 & male & 23 & \rv{Zhoukou} & I get nervous when I stutter and I will speak very slowly.\\ 
     9 & male & 24 & \rv{Fuzhou} & I stutter more on the important occasions.\\ 
     10 & male & 23 & \rv{Huhehaote} & Stuttering has negative effects on my career.\\ 
     11 & female & 22 & \rv{Shenzhen} & I stutter more when making phone calls or talking to strangers.\\ 
     12 & female & 22 & \rv{Yinganmeng} & I repeat, stretch and could not pronounce some words.\\ 
 \hline
\end{tabular}
}
\caption{\label{tab:table-name} Meta information and self-descriptions of the 12 formative study participants}
\Description{This table shows the meta information and self-descriptions of the formative study participants. We have 12 participants (9 male, 3 female, aged 17-35) with various stuttering situations.}
\label{table:participant}
\end{table*}

\subsection{Analysis of Content Posted in Chinese PwS Online Communities} \label{Observation}
We first searched for online communities that are frequently visited by Chinese PWS, who share their experiences and seek help from others, and identified the following ones: 
Weibo\footnote{Weibo: https://weibo.com/} \rv{(246 million daily active users (DAU))},
Zhihu\footnote{Zhihu: https://www.zhihu.com/} \rv{(30 million DAU)}, 
Baidu Tieba\footnote{Tieba: https://tieba.baidu.com/index.html}  \rv{(10 million DAU)},
and Douban\footnote{Douban: https://www.douban.com/} \rv{(3 million DAU)}.
We screened posts by the keywords ``Stutter Correction'', ``Stutter Treatment'', and ``Stutter Exercise'' from these forums. All acquired data is from 10 years prior to the collection date (June 10th, 2022). After data cleaning, 807 representative posts were eventually obtained. The data was then thematically analyzed using an open coding approach \cite{khandkar2009open}. \rv{Two co-authors read through the collected posts first to get familiar with the data. Then they coded the data independently by grouping similar posts and extracting high-level topic keywords. On the weekly project meetings, all co-authors discussed the coding result together and updated the code book.}

\subsection{Semi-structured Interview}
To validate the findings from the online posts analysis and to identify other potential user needs, we conducted online semi-structured interviews with 12 PwS (9 male, 3 female, aged 17-35) with a variety of self-described stuttering situations (See Table \ref{table:participant}). \rv{For participants screening, we referred to the criteria used by Sicotte et al. \cite{sicotte2003feasibility} that participants who were willing to participate should have demonstrated difficulty with at least 5\% of syllables spoken and considered themselves as having stuttering problems. To ensure the diversity of the participants, we selected participants from different locations in China. Each interview session took about one hour and all interviewees were compensated in accordance with local standards.} The interview focused on their stuttering experiences, workarounds, experiences of practicing speaking with others and expectations regarding assistive technologies. The development of interview questions on stuttering experiences was guided by Yaruss's evaluation of stuttering therapy \cite{yaruss2001evaluating}. Each interview lasted approximately 40 minutes. To analyze the interview data, we followed the Grounded Theory \cite{oktay2012grounded} method. \rv{Three co-authors read through the interview transcripts first to familiarize themselves with the data and then did the open-coding process independently. Then all co-authors discussed and updated the code book on the weekly project meeting for two weeks.} \st{We open-coded the interview transcripts and observations and then grouped them into four categories based on types of features PwS need: practicing scenarios with pressure, visual indicators like tone and volume, supportive community and mechanisms for persistent practice.}

% Key findings: we want to foreshadow the importance of the features we have in our prototype here with the findings: 
% practice scenarios that can induce different stress levels, deliberate practice (know the issues in current practice accordingly, receive timely personalized feedback to further improve), etc. 
\rv{\subsection{Findings}
In this section, we will report our findings from the formative study and highlight unique aspects of the Chinese context.

\subsubsection{Needs for speech practice apps for PwS}
We found that PwS in China suffer from shortage of SLP and efficient practice strategies. Therefore, they all desire assistive mobile applications that could help them improve speaking fluency. However, there are few studies or products designed for them. All these findings emphasised on the necessity of designing such practice assistance apps for PwS.

\paragraph{Shortage of SLP in China}
We were surprised to find that the attitudes of PwS towards SLP in China and in western countries are very different. As aforementioned, there are many works helping SLP monitor the progress of their stuttering patients in the western context \cite{madeira2018personalising, madeira2013building} and they reported PwS prefer feedback from SLP than other people \cite{McNaney2018stammerApp}. Although qualified SLP could assist and benefit PwS for their speech, SLP service is not always available in China due to its high cost and insufficiency. From the online posts, we found that it is difficult for PwS to find reliable SLPs due to the insufficiency of qualified SLPs in China (\textit{``Can anyone recommend reliable SLPs for me? It is hard to distinguish by myself''}). Meanwhile, there are many PwS accusing SLP and stuttering-treatment organizations for deceiving and over-charging them (\textit{``The characteristics of their therapy: hype concept, long duration and high costs''}). 

\paragraph{Practice speaking alone as the main strategy}
Considering this situation, we asked our interviewees about their current workarounds. We found that their primary method for improving speech fluency is to practice reading and speaking everyday. Speaking practice was perceived to be more efficient than reading (\textit{P4: ``I used to spend 30 minutes per day reading books aloud. Later I found that practice speaking is more efficient than reading''}). In addition, speaking practice could be done independently by talking to yourself or dependently with others' response. However, practicing alone is not efficient as practicing with others, as most PwS found speaking to themselves totally different from talking to other people (\textit{P1: ``I do not stutter at all when I am talking to myself''}). Unfortunately, it is hard to find appropriate practice partners that are patient and could provide constructive advice. 

\paragraph{Lack of assistive mobile applications designed for PwS}
Seeing the limitation of the SLP and difficulty of practicing with others, we asked our interviewees about their experience of using assistive technologies such as mobile applications for speech fluency practice. All interviewees reported that they never used mobile apps specifically designed for PwS. P2 used an app only for assessing general speech fluency and Mandarin pronunciation, with no specific stuttering-related functions. P4 expressed her concern about the credibility of mobile applications: \textit{``I want to learn more about speech training apps, but I am also worried of being deceived''}. When asked about applications specifically design for PwS to practice speaking fluently in simulated scenarios, all interviewees indicated that they were eager to try such applications but could not find any in the app store or other social media platforms in China. Similar findings were reported by McNaney et al. in their survey study \cite{McNaney2018stammerApp} that most respondents have not used any mobile apps designed for PwS. It is obviously a general problem with PwS worldwide.

\subsubsection{Stuttering situations}
\paragraph{Common and specific scenarios in China} 
We identified 8 different scenarios in two main categories: official and daily life situations. Some of them overlap with prior studies but some of them are unique to Chinese context. Our interviewees indicated that they stutter when delivering a speech \cite{ghai2021fluent} (P5, 9, 11), attending interviews (P1), presenting in a meeting (P4, 5, 7, 10), having workplace conversation  (P4, 5, 7, 9, 10), answering phone calls \cite{McNaney2018stammerApp} (P1-12), shopping (P2, 4), ordering food (P4) and making friends (P11). Meanwhile, there are also some specific scenarios for Chinese PwS. For instance, reporting identity information when doing community level PCR test (P2) and reporting code for getting delivery (P4). These scenarios usually cause people nervous so that they stutter more. This leads to the first design consideration \textbf{(DC 1): The practice should focus on the stutter-trigger scenarios, which is personalized across individuals and cultural context.}

\paragraph{Personalized nervous scenarios} 
Another interesting and unique finding is that some people got nervous when talking to their acquaintance (P4, 10, 12) while others stutter more when they are talking to strangers. Though there are personalized differences, it is common that people stutter more when there is some real persons around them than being alone, no matter whether those persons are talking to them. In addition, when comparing phone calls, video calls, and in-person interactions, the majority of our interviewees (N=11) considered in-person conversations to be the most difficult, expect that P6 was more afraid of phone calls since she has difficulty pronouncing the initial syllable of a dialogue. Not being able to allow the caller to see her face aggravated the problem, as she would have liked to signal that she is attempting to speak out. Based on these findings, we concluded \textbf{DC 2: The practice scenarios should be carefully designed to trigger pressure and nervousness.}

\paragraph{Hard-to-pronounce syllables}
Many PwS have difficulty in pronouncing some syllables such as `da' and `li'. This is different from English-speaking PwS as they stutter on words like `want' and `sam' \cite{ghai2021fluent}. And they are usually blocked on those syllables when they are nervous. In addition, P1 and P10 indicated that they have some persistent hard-to-pronounce syllables, which will block them whether they are nervous or not. Therefore, they think practicing different sentences containing the specific hard-to-pronounce syllables without scenario would also help. This derived \textbf{DC 3: The practice scenarios should require users to say stutter-trigger syllables, either implicitly or explicitly.}

\subsubsection{Personalized timely feedback and social support}
\paragraph{Personalized peer feedback and advice}
As aforementioned, PwS in China rely more on some types of peer support than SLP support. Although there are many online forums and communities, it is still difficult for PwS to get support and the answer they need in time. In every popular online forum, we observed a lot of posts where PwS describe their situations and seek for peer suggestions or recommendations for potential methods to improve their situation (\textit{``I have been pronouncing like this for decades and I really want to change. Does anyone have potential methods?''}). However, the response rate is low. The reason may be that many users just enter the forum to search for the information they need and do not look at other posts. Many PwS have tried a lot of methods learned online, but they found it hard to persist for a long time due to lack of personalized advice and clear vision of their progress. Advice in the online forums are just generic ones while no specific personalized feedback based on their performance were provided. So they don't know what is their exact speaking problem and the most suitable and effective methods (\textit{``I tried many methods such as metronome and deliberate breaks, but I quit after a short time because I did not see any progress''}). When asked about the potential effect of the personalized feedback from other PwS, all interviewees agreed that it will benefit their speech fluency improvement and give them mental support and motivation (\textit{P9: ``With a supportive group, I will be more motivated to practice because I know that they are practicing with me and will give me feedback''}). Many interviewees also emphasised on the importance of finding supportive peers (\textit{P4: ``I would be more confident if I receive positive feedback from other PwS'' and P5: ``I want more personalized authentic feedback from others, not just common compliments''}). Therefore, instead of providing connection with SLP, feedback and support from other PwS should be valued more in our design - \textbf{DC 4: The platform should build a supportive community where PwS could give and receive personalized peer feedback in time.}
% feeling of practice together?
% PwS agreed that they find it difficult to identify their speaking issue and they are keen to gain feedback and advice from other PwS. In the meantime, they also want to see others sharing their experiences of fighting against stuttering using copying strategies.

\begin{figure*}[h]
    \centering
    \includegraphics[width=\textwidth]{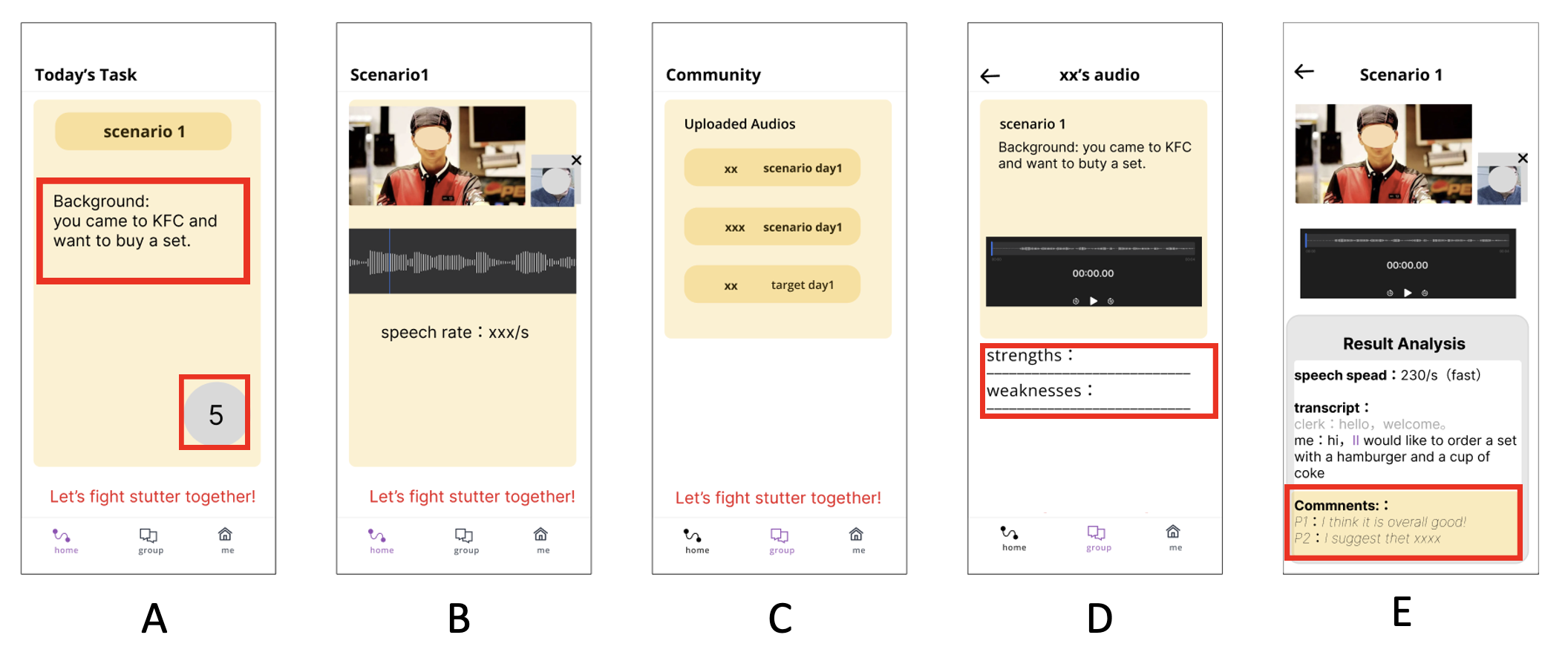}
    \caption{Low-Fidelity Prototype UI Created with Figma (A. Task description page with a 5-second countdown before starting practice. B. Practice page with real-time objective feedback including facial expression, volume and speech rate. C. Community page where users give and receive feedback within the Community. D. The review page where commenters can view the task background, listen to the audio, and comment on the strengths and weaknesses. E. Review page where the practicer can view others' comments)}
    \label{fig:lowfi}
    \Description{This figure displays the low Fidelity Prototype UI. Five main pages are displayed from left to right: A. Task description page with 5-second countdown before starting practice (with red rectangle annotation around the paragraph of task description and countdown). B. Practice page with real-time objective feedback including facial expression, volume and speech rate. C. Community page where users can listen to others' audios in the community. D. The review page where commenters can view the task background, listen to the audio, comment on the strengths and weaknesses. E. Review page where the practicer can view others' comments.}
\end{figure*}

\paragraph{Timely objective feedback}
When asked about the detailed types of feedback they want, many interviewees also mentioned the lack of quantitative and objective indicators (\textit{P1: ``When I practice for a speech, I do not have any objective feedback. I can just receive some simple suggestions from my friends'' and P3: ``I assessed my behaviors based on my own subjective feelings, without any feedback from other aspects''}). When asked about the detailed types of feedback, many PwS (N = 7) mentioned speech rate, tone and transcript, which aligns with the findings from prior work on providing instant feedback on analysis and training of general speech fluency \cite{fung2015roc, zhao2017semi, engwall2004design}. Although these features were proved to be beneficial for general speech training, their effectiveness in assisting PwS practice speaking fluency was not investigated. This is also evidenced in the online posts as many of them indicated that PwS should try to speak slowly to avoid stuttering. In addition, many PwS consider facial expression and body movements as important factors to pay attention to when they are speaking, because their facial expression and body movements are somewhat influenced when stuttering and they want to avoid that (\textit{P7: ``The most recent practice experience of mine is to have video conferences with my friends, where I could clearly see my facial expressions''}). This is also reflected in the forum posts (\textit{``Paying attention to your facial expression is important! Some PwS have strange facial expressions even when they are not stuttering''}). Therefore, real-time self-image including facial expression and body movement should be considered as an important type of feedback for PwS to adjust accordingly - \textbf{DC 5: The platform should provide timely objective feedback for PwS to reflect on and adjust accordingly (e.g. speech rate and volume).}

}

\section{Prototype Design} \label{interfaceDesign}
With the guidance of the five DCs, we designed several features and integrated them into a low-fidelity prototype  \cite{Rettig1994Prototyping} to elicit more honest feedback on the effectiveness of these key features. \rv{By conducting an iterative design workshop with the 12 participants from the formative study, we summarized the key findings for potential improvements on the prototype and then polished the prototype accordingly to the final version.}

\subsection{Low fidelity Prototype}
We created the low-fidelity prototype (See Fig \ref{fig:lowfi}) using \textit{Figma}\footnote{Figma: https://www.figma.com/}. \rv{In the following paragraphs, we present the key features of the prototype in two perspectives: the practicer and the commenter.}

\rv{\subsubsection{Practitioner's perspective - Fig \ref{fig:lowfi} A, B, E}}
When starting a task, a paragraph of the task description page with a five-second countdown will be displayed before the conversation task officially starts (Fig \ref{fig:lowfi} A). The countdown is designed to give users a moderate sense of tension \textbf{(DC3)}. After the countdown, the practice page will be loaded where a video will be auto-played (Fig \ref{fig:lowfi} B). The content of the video is a person talking to the user, prompting questions and waiting for users to answer. We recorded the video of a real person talking instead of using animations to make it more stressful, as participants reported they become nervous when there are real persons around \textbf{(DC3)}. The person in the video will nod and smile to make the user experience more realistic and stressful \textbf{(DC3)}. \rv{For conversation topics, we used the ones mentioned by our PwS interviewees which could induce stutter-triggering words \textbf{(DC2)} and make them nervous, including job interviews, phone calls, workplace meetings, clothing purchases, food orders, and campaigns \textbf{(DC1)}. Some of them were also mentioned in the online survey conducted by McNaney et al. but not implemented in their prototype (StammerApp) \cite{McNaney2018stammerApp}. In addition, some specific pronunciation tasks were also designed where users need to read the text displayed aloud \textbf{(DC2)}. The text was designed to contain several stutter-triggering syllables. For this kind of task, the video did not include any real person and only displayed plain text. }

While the user is speaking, some real-time objective feedback \textbf{(DC5)} will be displayed below the video: a dynamic volume graph and speech rate number (Fig \ref{fig:lowfi} B). A real-time self-image captured from the front camera will be displayed at the bottom-right corner of the video area. From the self-image, PwS can observe their facial expressions and body gestures when speaking. We expect the users to reflect on these objective feedback indicators and adjust their speech accordingly in real-time. The users' voices will be recorded and uploaded to the server for the next phase. In addition, on all the main pages, a paragraph of encouraging words (See the text in read in Sig \ref{fig:lowfi}) will be displayed at the bottom of the page \textbf{(DC4)}.

\rv{
Besides real-time self-reflection and adjustment based on the objective feedback, users can also reflect on their behaviors after practice by opening the tasks they completed to view others' feedback on their own audios (Fig \ref{fig:lowfi} E). The result page displays the recorded audio, average speech rate, audio transcript and comments from other PwS users (commenters). All comments will be evaluated by researchers to verify that none are toxic (\textbf{DC4}). Although prior work also allows users to review and reflect, they can only listen to their audios, without any indications or feedback \cite{McNaney2018stammerApp}.}
% Here we have the UI for daily report but did not report due to space limit.
% In addition, encouraging words are displayed in every main page in the prototype (text in red in Fig \ref{fig:lowfi} E).

\subsubsection{Commenter's perspective - Fig \ref{fig:lowfi} C \& D}
After finishing the task, the user can check audios uploaded by other PwS users in the community (Fig \ref{fig:lowfi} C). By opening one audio tab, a page with detailed information will be loaded (Fig \ref{fig:lowfi} D) where users can see the task description, listen to others' audio and give them comments and grades as feedback on the practicer's strengths and weaknesses \textbf{(DC4)}. Everyday, all users will be reminded by Wechat\footnote{Wechat: a Chinese instant messaging, social media, and mobile payment app developed by Tencent. https://www.wechat.com/} private messages to practice, give feedback to other peers and check the feedback they got in time\textbf{(DC4)}.

\subsection{Iterative Design Workshop with PwS}
To collect primary user feedback and perception on the prototype, we recorded a video to demonstrate the workflow and interaction in this low-fidelity prototype. Then we invited the 12 interviewees back to experience our low-fidelity prototype and collected their feedback by interview. Each session took about 30 minutes. In the following paragraphs, we present the findings from the iterative design workshop on user experience and conclude improvements to be made for each feature.

\rv{
\subsubsection{Insufficient Stress of the tasks}
All participants found the personalized simulated video conversation tasks beneficial as they could practice and rehearse some stutter-triggering scenarios. However, many participants (N=8) mentioned that the pressure in the task was insufficient to train them to handle similar circumstances under pressure in the real world. In terms of potential solutions, P4 suggested that character in the task video should make direct eye contact with users (\textit{P4: ``People in the video should have direct eye contact with users to make it more stressful''}). P5 proposed hiding the task description after a few seconds to make the simulation more realistic and demanding. Many participants (N=9) also felt that situations requiring prompt response are stressful. In addition, all task themes should be relevant to daily life; otherwise, PwS may find it difficult to complete the tasks and become frustrated (P6). Although most participants required higher stress level for the tasks, P6 stated that different stress levels should be provided for each task, so that everyone could choose levels suitable for their practice. This aligns with our findings in the formative study that the preferences for tasks of PwS are highly personalized. 

Based on the participants' feedback, we iterated the prototype by \textbf{(1) asking the persons recording the video to make direct eye contact with potential users, (2) hiding the task description after a few seconds, (3) recording task videos that require prompt response and (4) providing multiple stress levels of each task by adding disruptions.}

\subsubsection{Number of conversation rounds in a task}
Most participants (N=10) appreciated the conversation content in the video as it is relevant to daily life. However, many participants (N=7) found the task insufficiently engaging and stressful because the persons in the videos only spoke once and then waited for the user to speak. When the user has finished speaking, the person in the video should speak up again to make the dialogue sound more natural. Therefore, \textbf{we increased the number of conversation rounds in a task in the iterated prototype.}

\subsubsection{More personalized real-time feedback and subjective feedback}
All participants found the visualization of speech-related data helpful, including speech rate(P1, 2) and facial expression (P2, 7). In addition, they also desired to see more types of speech indicators in real-time. Meanwhile, they have different preferences for the types of real-time feedback, such as volume (P6), pitch, tone, rhythm (P1) and transcript (P3). To fulfill the personalized needs of real-time feedback, \textbf{all forms of real-time feedback indicators mentioned by the participants were presented by default in the iterated prototype. And switches for each indicator were provided for users to switch off whatever indicator they did not want to see in real-time.}

All participants believed the subjective feedback would be beneficial and they were willing to carefully listen to and provide comments on the audios of others. Meanwhile, P2 and P8 expressed their concern about the situations that they do not have any concrete suggestions to offer. To address this problem, we decided to \textbf{add a rating function as one type of compulsory feedback and to make comments optional,} so that users can only provide ratings if they have no comments.

\subsubsection{Insufficient reflection assistive functions}
Most of the participants (N=9) agreed that the current design of the reflection page was beneficial. In addition to the reminders and encouraging words, P5 and P7 suggested that a daily report demonstrating progress for the day could help them better reflect on their daily behaviors and progress, which could be encouraging and motivating for future practice. Therefore, we decided to \textbf{add a daily report function.}

}

\subsection{Final Prototype}
We modified the prototype based on the findings from the iterative design workshop. In this section, we present the final version of our prototype, including the practicer's perspective and the commenter's perspective.

\begin{figure*}[h]
    \centering
    \includegraphics[width=\textwidth]{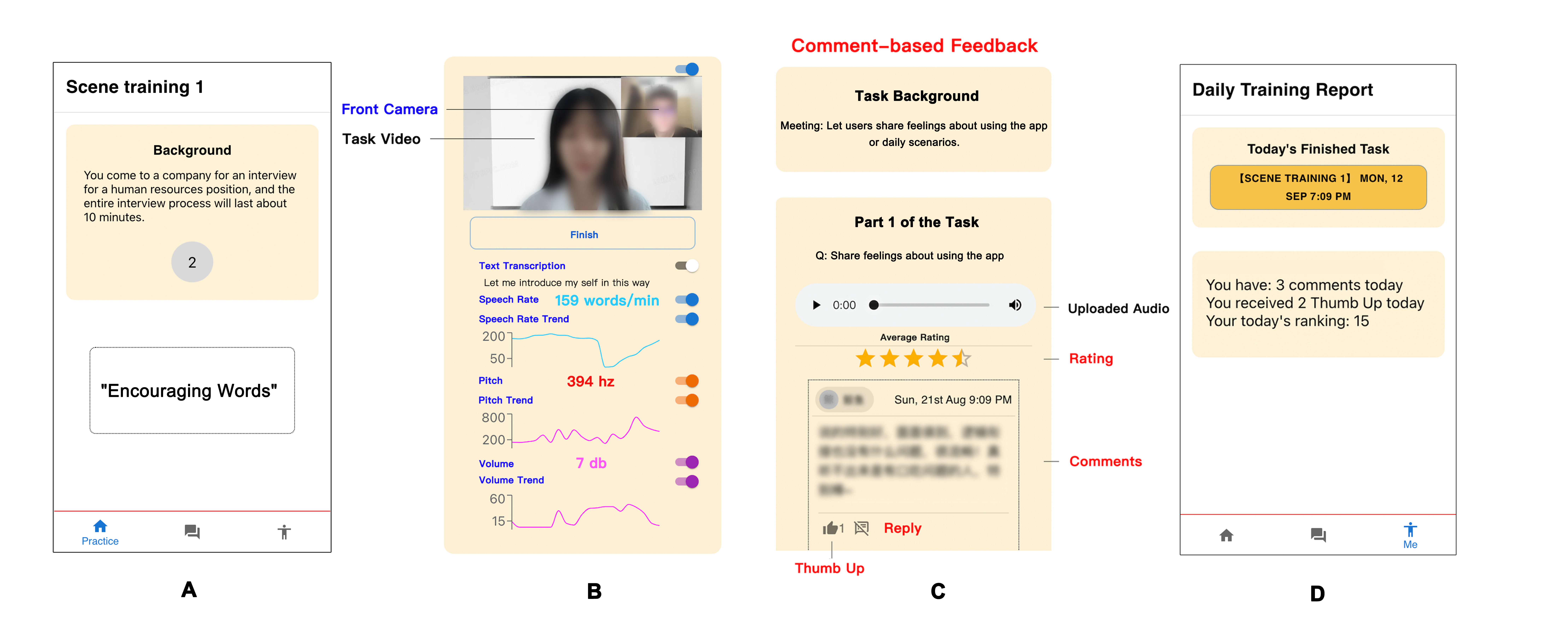}
    \caption{The workflow of practicers' perspective: task description page (A), practice (B), review (C) and daily report (D)}
    \label{fig:phase1}
    \Description{The workflow of practicers' perspective: (from left to right) task description page, practice page(with real-time feedback indicators), review page and daily report.}
\end{figure*}

\subsubsection{Practicer's perspective} \label{finalPracticer}
Practicer will first see the task description page (Fig \ref{fig:phase1} A) with text description and a five-second countdown, which is the same as the low fidelity prototype. After the countdown, the practice page (Fig \ref{fig:phase1} B) will be loaded, where the conversation task video will be auto-played and users need to talk to the person in the video. The person in the video will nod, smile, have direct eye contact with the users and interrupt them during the conversation by saying words like ``excuse me'' to make the user experience more realistic and stressful. 

While the user is speaking, a set of real-time feedback will be displayed below the video: audio transcription, speech rate, pitch, volume and their trends accordingly (Fig \ref{fig:phase1} B). A real-time self-image captured from the front camera will be displayed at the top-right corner of the video area, From which PwS can observe their facial expressions when speaking. All the real-time feedback can be individually switched off by the corresponding buttons to their right. There are several sections in each video. In each section, the person in the video will say several sentences to set up and push forward the scenario, ask a question and then wait for the user to respond. This is to increase the stress level by increasing the conversation rounds. Upon finishing each section, users need to tap the "Finish" button to indicate that they finished speaking. All users will be reminded by Wechat private messages to practice every day.

By the end of each day, users are reminded to check the feedback they get from other users (\textbf{DC4}) and the daily report. By opening the review page of a task, users can re-listen to their audio and see other users' comments (Fig \ref{fig:phase1} C). The daily report concluded the tasks the user finished today (Fig \ref{fig:phase1} D), which could be opened to see others' feedback, the number of comments this user gave to others, the number of thumb-ups received and daily ranking.

% The reason of not detecting the dialogue completion automatically is to avoid mis-detection due to stuttering. 

\begin{figure*}[h]
    \centering
    \includegraphics[width=0.65\textwidth]{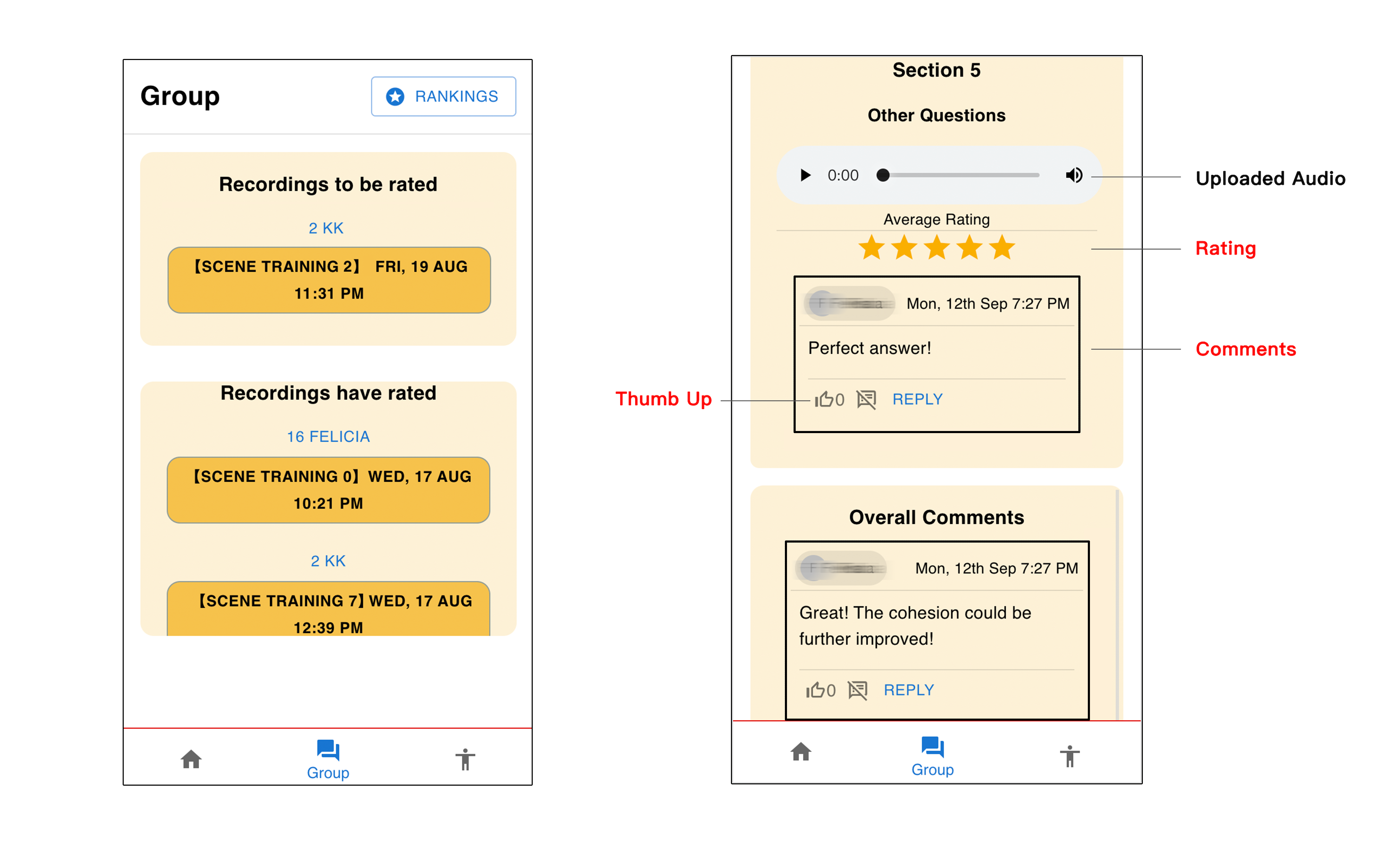}
    \caption{The workflow of commenters' perspective: group page (left) and comment page (right)}
    \label{fig:phase2}
    \Description{The workflow of commenters' perspective: group page (left) where commenters' could view others' trials and comment page (right) where commenters need to listen to the audio and fill in the text area with their comments.}
\end{figure*}

\subsubsection{Commenter's perspective} 
After finishing all compulsory tasks, users are required to open the community page (Fig \ref{fig:phase2}, left) to check others' audios. By selecting an audio block, a page with detailed information will be loaded (Fig \ref{fig:phase2}, right) where users can see and listen to others' audio and give them feedback. For each section of the audio, users need to rate the audio (from 0 to 5 stars) and give optional comments. Users will be reminded to point out the issue in a polite manner and provide practical advice with encouragement. Users can also thumb up and reply to others' comments (See the red labels in Fig \ref{fig:phase2}, right). All comments are evaluated by researchers to verify that none are toxic.

\subsubsection{Prototype Implementation}
We utilized React\cite{gackenheimer2015introduction} framework for the front end and Node.js for the back end. For the database, we used MySQL and used Structured Query Language (SQL). For real-time features, We utilize the real-time speech-to-text conversion service from Tencent Cloud\footnote{Tencent Cloud: https://www.tencentcloud.com/}, a world-known back-end service provider. The service packs the blob data from each audio frame to the server and gives text feedback in real time. By using \textit{getUserMedia} method in a web application, we can get control of the front camera and recording system of the device, and continuously get the underlying logic \& resources of video and audio. Furthermore, we use Fourier analysis and signal processing to extract volume and pitch information and display the real-time image from the front camera.

\section{Deployment Study} \label{evaluaiton}
We conducted a seven-day deployment study with multi-user online simultaneously to understand the effectiveness of the key features we designed, and explore potential design implications for future work. \rv{To our knowledge, it is the first time such a prototype was designed and evaluated in a long-term, multi-user simultaneous online deployment study.} 

\subsection{Participants and Apparatus}
We recruited 11 participants (8 males, 3 females, aged 19-35) from the online community by posting recruitment messages. Considering that some PwS do not want to emphasize their stuttering and tend to live with it life-long without any training, we only recruited those who were willing to improve their speech fluency using our prototype. During registration, we asked the participants to rate the severity of their stuttering on a scale from 1 to 10. The results ranged from 4 to 10. We deployed the prototype to a remote server so that participants could use it to practice from anywhere and on any devices convenient for them, as long as they had access to the internet to open the website. Participants were compensated in accordance with local standards.

\subsection{Task Design}
Every day, each participant was assigned three tasks: one compulsory conversation task, one compulsory pronunciation task, and one optional conversation task. Each conversation task took about 10 minutes and each pronunciation task took approximately one minute. Only after finishing all required tasks could the user view and comment on the audio of others. Each participant was asked to comment on three assigned participants each day, which was fixed throughout the experiment. This was to ensure each participant received feedback from others and commenters could observe potential improvements in the participant they were evaluating.

\subsection{Procedure}
To begin, we held 11 individual onboarding sessions one day before the official commencement day of the study, where participants were briefed about the study and signed consent forms. Then participants watched a pre-recorded video demonstrating the prototype and explored the prototype freely with their assigned account in order to familiarise themselves with it. Participants were able to ask any questions about the study and the prototype. After that, participants were asked to finish a series of tasks that walked them through the entire daily practice procedure: talk to the video, listen to and comment on others' uploaded audios and review the comments for their own audios.

For the following 7 days, participants were required to complete the assigned tasks and comment on others' audios in a timely and appropriate manner every day. At the end of a day, they were asked to fill in a daily questionnaire to reflect on their experience for the day, where they need to reflect on their training and rate each type of feedback for their usefulness from 1 to 7. On the day following the seven-day study, All participants attended a semi-structured interview, where they were asked about subjective reflections on the prototype and expectations for future systems. As aforementioned in Section \ref{interfaceDesign}, all users were reminded by WeChat private messages to practice and give comments.

\subsection{Data Analysis}
Three of our co-authors conducted an open coding process to analyze the qualitative data, including user comments, responses to the daily questionnaires and interview transcripts. In addition, we collected activity logs of users' behavior in each task, such as the audio length, status of each indicator switch, and comment-related statistics. After the data cleaning process, we ran statistical tests on all the quantitative data collected.
\section{Results}
\rv{In the following paragraphs, we will discuss the findings of user perception and usage of the personalized practice scenarios, timely targeted feedback and social support for PwS (RQ2).}
\st{Here, we present our findings of the deployment study. The majority of participants completed their tasks on time and provided constructive feedback to others. Consistent with what we discovered from the literature review and formative study, the usage patterns of the individuals vary, but there are also many similarities. In the following paragraphs, we will discuss the findings based on the four Design Considerations derived from the formative interview: (1) Practice Tasks in Different Scenarios with Pressure; (2) Real-time feedback; (3) Comment-based feedback from community; (4) Motivating methods.}

\subsection{Practice Tasks in Different Scenarios with Pressure}
Many participants indicated that the tasks are useful and engaging for them to practice speech fluency. Meanwhile, there is still room for development in the task's design in order to make it more personalised and realistic.

\subsubsection{Overall Usage of the Tasks}
Overall, all participants indicated that the tasks are helpful for them to practice speech fluency in an effective and engaging way. In addition, all participants stuttered several times in the practice procedure, and they agreed that the task was stressful and embedded stutter-trigger words, which satisfied their practice requirements. Specifically, the tasks trained them to organize the speech content quickly before speaking (\textit{P11: ``I find myself organizing the speech content better and faster. I also have more courage to speak in real life''}) and speak for a long time (\textit{P5: ``In real life, I don't have opportunity to speak for a long time''}). P3 expressed his excitement about finding a potential reason for his stuttering: \textit{``Through the tasks, I found that my stuttering may not be physical defect, but caused by my inability of organizing the speech content quickly. I'm more motivated to practice''}. In addition, many participants (N=5) indicated that many tasks were very engaging and made them feel nervous to some extent, which is helpful because it simulated the situation in which they need to talk under pressure in real life (\textit{P2: ``Many tasks made me nervous and under-pressure, which I believe could help me prepare before facing similar situations in real life''} and \textit{P6: ``I found it very helpful because I can't find people to practice speaking in a simulated situation. It helped me to train speaking for daily life and workplace scenarios, which is just what I need''}).

\subsubsection{\rv{Personalized Tasks \textbf{(DC1)} with Stutter-triggering Syllables \textbf{(DC2)} } }

\paragraph{Preference for task sections} \rv{Overall, all participants agreed that the tasks were designed properly and well simulated the conversation scenarios in real-life.} Our results show that the number of sections in the task video and task topic influenced the user's experience. As aforementioned in Section \ref{interfaceDesign}, we designed the tasks to contain various number of sections from 1 to 5. Many participants (N=6) indicated that they prefer tasks with at least five sections over those with only one or two sections that require them to deliver a lengthy speech (\textit{P1: ``Tasks with 5 sections resembled conversations in daily lives, which are the situations that I need to practice the most''}). Other participants preferred tasks with few sections (\textit{P10: ``I prefer the lengthy speech tasks because I tend to stutter more when speaking for a long time without interruption''}).

\rv{\paragraph{Preference for task topics}}
We also asked each participant to identify their favourite tasks. It was found that their preferences varied a lot and they wish to receive more personalized tasks in the future. As mentioned before in section \ref{interfaceDesign}, our simulated conversation tasks included two types: daily life scenarios and important occasions like in workplace. Seven participants preferred the important occasions (\textit{P6: ``I like the meeting in the workplace, because it is important for me to practice''}) and the other four preferred daily life scenarios (\textit{P5: ``I like the movie sharing task because it's interesting so I can say more''}). From the daily questionnaires and interviews, we found that the preference for tasks was highly related to their personalized needs in their daily lives. For instance, there is one task that required participants to deliver a speech for campaign in the university. Only P5 and P10 liked this task as they are undergraduate students, and it can help them prepare for the campaign in real university life (\textit{P10: ``This task is a situation I normally worry about and stutter extremely severely.''}). However, other participants indicated that they dislike this task the most, because it is too far away from their daily lives (\textit{P5: ``I know little about this topic, so I don't know what to say''}).

\subsubsection{Stress and Realism \textbf{(DC3)}}
As mentioned before in section \ref{interfaceDesign}, we designed and recorded task videos with real persons who performed verbal interruptions, direct eye contact and nodding to make it more realistic and stressful. Many participants (N=7) indicated the videos were well designed and made them nervous to some extent (\textit{P7: ``At first I believed that I was talking to a real person in real time because the person in the video keeps nodding and smiling'''}). We also observed different preferences toward stress level among all participants: most participants preferred high stress levels while there were also some participants preferred low stress levels (\textit{P6: ``I suddenly did not know what to say when she interrupted me. It feels awful and I would rather not practice that kind of scenario recently''}). However, some participants also stated that there is still a gap between practicing with the prototype and the real world. P11 said that the conversation design is not realistic enough: \textit{``The questions the waiter asked in the ordering task are not similar to what I heard in real life, which made it less realistic''}. P9 mentioned the problem that the actors in the task videos could not bring enough pressure: \textit{``At first, I'm very nervous. But as the study went on, I got familiar with it. And the actors are not like my boss, who gave me a lot of pressure''}. From the audios uploaded, we discovered that some participants did not treat the tasks seriously as real-life scenarios. They kept silent for a long time before speaking for some tasks, which is not likely to happen in real life, if they have no idea (P11: \textit{``I really don't know what to speak for some unfamiliar topics, so I need to think for a while before speaking. If this happens in real life, I may just tell the listener that I have no idea''}). They quit in half of this task when they thought their performance is too bad and re-open it to start again, while in real life, they do not have the chance to do that (P11: \textit{``I just do not want others to hear me speaking so bad''}). 

\subsection{\rv{Personalized Timely Objective and Subjective Feedback} 
\sout{Real-time Objective Feedback}
}
\rv{
In this section, we present the participants' perception of comment-based subjective feedback and real-time objective feedback.}
\sout{
In this section, we present the finding of participants' usage of real-time (\textbf{DC2}) feedback. Overall, timely comment-based feedback from a supportive community were valued more than real-time objective feedback.}

\subsubsection{Timely Subjective Peer Feedback and Social Support \textbf{(DC4)}}
\paragraph{Overall usefulness of the subjective peer feedback}
As aforementioned in Sec \ref{interfaceDesign}, we designed four features for peer feedback: text commenting, rating, thumb-up and reply. In the seven-day deployment study, we collected 1676 valid text comments in total, 47 of which were replied and 45 of which got thumb-up. For comment-based features (comments (M=4.48, SD=2.73), thumb-up (M=2.24, SD=1.89) and rating (M=2.03, SD=2.38)), we conducted a one-way repeated measures ANOVA test to determine the difference between each factor. The result showed that the differences of participants' opinions between comments and thumb-up (p=0.004), comments and rating (p=0.014) are significant, which indicated that comments were the most important one in the set of comment-based features.
\rv{
After open-coding, all 1798 comments are concluded into three main categories according to their content: problems and advice(1059), compliments and encouragement(550) and reply to the content (80). Some comments may fall into more than one category. Overall, participants found most of others' comments beneficial for diagnosing their problems, providing advice, encouraging them to keep practicing, and communication with others. In the following paragraphs, we will discuss user perceptions of each type of comments in detail.

\paragraph{Problems and advice aided PwS to diagnose their problems}
Problems and advice: The majority of the comments pointed out specific issues and gave personalized advice about speech rate(361), cohesion(283), content(169), emotion and confidence(153), clarity(68), volume(29) and accent(6). All participants found others' advice generally helpful, although there were also some disagreements with some comments.}
For speech rate, we found two different speech rate groups often comment on each other suggesting them to change their speech rate (\textit{P1: ``try to rise your speech rate steadily'' and P2: ``You speak too fast. Try to slow down and take a break at the appropriate position, which will make you sound more like a normal person''}). Although four participants found speech rate related comments helpful (\textit{P9: ``People advised me to talk more slowly. I believe it will help''}), P4 expressed his disagreement with the comments: \textit{``Comments told me to speak slowly but I don't think it is a problem''}. Another issue frequently mentioned in the comments is speech cohesion, including abuse of filler words (\textit{P5: ``You used too many words like `umm', `well' and `you know'''}) and unnecessary blocks in one sentence, which may be caused by improper breathing pattern and problems with the content organization (\textit{P1: ``You can speed up your speech and reduce the number of pauses within and between sentences'' and P9: ``Think about it before saying it; stop a little bit in the middle of each sentence to give yourself time to think''}). For speech content, some comments suggested the participant speak more (\textit{P1: ``The overall quality is great but the content is short. I hope you can say a little more''}) while others suggested the participant speak less (\textit{P9: ``I think you can say less''}). Regarding mindsets and emotion, many comments suggested the receiver to relax and be more confident. Some participants also pointed out that some users should make their speech more emotional (\textit{P6: ``We should speak with emotion in life, like joy and sadness.''}).

When asked about how they gave comments to others, most participants indicated that they tried to point out problems gently and recommend useful methods they have tried, which is consistent with our hypothesis (\textit{P1: ``My expectation is to point out the problem to others, but it may not be accepted by everyone'', P2: ``My comments were all about speech therapy because I have learned and tried many speech therapy methods''}). P11 expressed his concern that giving comments may disturb others: \textit{``I was afraid that my suggestions will disturb the original methods they are using''}.

\paragraph{Compliments and encouragement helped PwS stay positive}
All participants were pleased and appreciative of others' compliments and encouragements. Many comments (486 out of 1798) were compliments about several aspects, such as speech rate and tone (\textit{P2: ``The speech rate is perfect'' and P3: ``I rated it 5 out of 5 because there were proper pauses and the speech rate was proper''}). There were also encouragements and reassuring phrases (N=64): \textit{P9: ``I believe you will become more fluent. Just devote more time to practice'' and P10: ``You are only 20 years old. You have a lot of potential and chances''}. All participants agreed that they were encouraged and motivated to keep practicing (\textit{P7: ``The encouragement made me more motivated to practice''}). However, many participants (N=6) also stated that they dislike comments that are too vague or seem perfunctory (\textit{P6: ``I did not say well but the comment said it is good''}). 

\paragraph{Reply to the content spoken facilitated communication}
In addition to comments on speaking fluency, there were also 80 comments replying to what the speech content of the uploaded audios: \textit{P6: ``Though I do not like group gatherings, I will still go to it since it is essential in the working place'' and P8: ``I have not heard of this song. I will search for it later''}. Many participants found it interesting and had a sense of community (\textit{P4: ``It was very interesting. I felt like talking to a friend''}).

\paragraph{Thumb-up, rating and reply to comments enriched the comments and communication}
Most participants found thumb-up and ratings to be much less effective than textual comments. In general, these two components are underutilized, and participants thought it less clear than the user comments (\textit{P1: ``Thumb-up is not so intuitive as direct comments'' and P3: ``When rating others: I don't want to give negative impact, so it may not reflect the true situation''}). In the meantime, we discovered that the content being thumbed up are mainly in four types: longer-than-average text; encouragement; personalized advice; and interesting content and conversations. This indicated that these types of comments are more beneficial for PwS than others.

Five types of replies to comments were observed from the coding result: (1) Expression for thanks: \textit{P10: ``Thanks for your comments. Let's keep practicing together'' and P7: ``Thanks for your encouragement and advice''}. (2) Disagreement, especially on the speech rate: \textit{P1: ``I think my problem is not caused by the speech rate as I stutter even when I speak slowly. It is true that I am more likely to stutter when speaking fast, but I think the biggest effect of speaking fast is that it may cause my speaking the wrong word instead of stuttering''}. (3) Confusion: \textit{P7: ``Sorry, I do not understand your comment''}. (4) Reply to the speech content, but not relevant with the stuttering situations: \textit{P9: ``I am in a small city, which is far from downtown'' and P10: ``There is always chicken in the countryside''}. (5) Agree and Add-on to other participants' comments: \textit{P6: ``You are right. I have a lot of syllables hard to pronounce'' and P10: ``Yes, it is hard for me to pronounce words starting with `b'''}. This indicated that these types of comments are more likely to trigger communication among PwS in the community.

\paragraph{Encouraging words on every page}
As aforementioned in Section \ref{interfaceDesign}, on each main page of \sys{}, reassuring phrases were presented to reassure and encourage users to speak with confidence. Most participants (8 out of 11) considered it encouraging whereas the remaining participants did not pay attention to it (\textit{P8: ``I didn't see it because I didn't need it''}). P2 even provided us with a collection of books and encouraging phrases. Besides encouraging words, participants indicated they want more interactive, intimate and practical encouragement. P1 wanted more interesting functions like sending roses to others and P4 recommended adding an additional screening condition (\textit{``Encouraging words can be displayed only when the average score in this task is lower than 3''}). P10 wanted brief and encouraging videos similar to \textit{``the King's speech''}. P8 suggested that human-recorded voice would be more encouraging than plain text (\textit{``I want encouragement with the real human voice, which could cause proximity and authenticity''}).

\paragraph{The daily report helped users reflect on their behaviors}
Although we expected that the daily report could help the participants see their progress clearly and reflect on their behaviors every day, its effect was limited. Most participants (N=6) did not open the report every day because they thought the tasks were the most helpful while the report was merely a summary of the information they already knew (\textit{P7: ``I seldom open daily report''}). Several participants (N=4) said the report was clear and facilitated their review of daily behaviors (\textit{P11: ``It is convenient for me to review quantitative feedback. With the report, I do not have to check every comment one by one''}). P1 stated that seeing the ranking in the daily report motivated him to practice more and comment more in the first few days of the study: \textit{``When I saw the ranking in the daily report, I was eager to get the first position in the ranking within the first four days. But later I did not know what to comment, as these problems were discussed in the preceding days''}.

\subsubsection{Real-time Objective Feedback \textbf{(DC5)}} \label{quant}
\rv{In this section, we present the findings of participants' perceptions of real-time feedback. Overall, all participants found the speech-related indicators displayed in real-time helped them gain a better understanding of their speaking, such as speech rate and stutter frequency. Different participants had their preferences for each of the real-time feedback metrics. They generally agreed more on the effectiveness of speech rate, audio transcription, and facial expression, while most of them did not consider pitch, volume, and their trend to be effective. In the following paragraphs, we will report the usage and user perception of each type of objective feedback}

\paragraph{The speech rate value was more straightforward than the rate trend and requires less cognitive load}
\label{sec: res-obj-Observation-speech-rate}
\rv{Speech rate is the second most useful feature (M=5.00, SD=1.71) according to the summary questionnaire and exit interview. Participants could directly know whether they are speaking too fast or slow from the speech rate value (\textit{P1: ``Speech Rate is very intuitive''}).} Some participants suggested that they would prefer a straightforward qualitative result of the comparison between their own speech rate and the optimal rate for Chinese speeches (150-180 syllables per minute \cite{li2010coping}), rather than exact values/the whole trend (\textit{P3: ``I could not look at so many indicators'' and P9: ``I just want to know whether I'm talking fast or slow''}). In addition, P1 mentioned that he wanted direct reminders to remind him when he speaks too fast or too slow rather than only displaying the speech rate value. Only one participant thought the speech rate did not help (\textit{P2: ``I can control my speech rate. I don't need to see the number''}) and he wanted a real-time auto-grading function based on speech rate and fluency, just like some Karaoke mobile apps.

\paragraph{Audio transcription displayed the stuttered words}
\label{sec: res-obj-Observation-text-transcription}
\sout{
The participants gave the audio transcription function a third high score. Although the average score for audio transcription was relatively high, the error bar also had a wide range, suggesting that this feature is still perceived very differently among participants (\textit{P1: ``From the transcript, you can tell if your speech is clear'', P2: ``Because I know exactly what I'm talking about, I don't know why it transcribed my voice into text'', P9: ``I looked at the transcript each time I spoke to see if there were any repeats or words not clearly pronounced'' and P11: ``Because I speak in dialect, the transcript may not be very accurate''}).
}
\rv{Audio transcription is the third most useful feature (M=4.91, SD=1.68). This feature is perceived very differently among participants. Some participants found it helpful in displaying the repetitions and words not pronounced clearly (\textit{P9: ``I looked at the transcript each time I spoke to see if there were any repeats or words not clearly pronounced''}). However, the effectiveness of these features is limited by the transcription accuracy (\textit{P11: ``Because I speak in dialect, the transcript may not be very accurate''}). In addition, there are also some participants finding it meaningless (\textit{P2: ``Because I know exactly what I'm talking about, I don't know why it transcribed my voice into the text''}).}

\paragraph{Facial expression caused pressure and needed to be paid attention to}
\rv{Facial expression captured from the front camera is the most useful feature (M=5.27, SD=1.81). Most participants found it useful to monitor their facial expressions and movements through the front camera view, so that they would feel more stressed or could adjust their facial expressions accordingly (\textit{while P5: ``The front camera helped me see my facial expression and mouth shape changes, and it caused more pressure'' and P6: ``It helped me a lot! I can see my facial expression all the time''}). While there was also one participant who found it not effective, as he did not care about his facial expression and only cared about his speech content (\textit{P2: ``I think it is optional''}).}
\sout{For the facial expression in the real-time self-image from the front camera, the standard deviation across participants (Figure~\ref{fig:daily-avg}) is ranked at the top1, which indicates the extremely controversial conditions. This view was corroborated in the follow-up interview: \textit{P2: ``I think it is optional'' and P5: ``The front camera helped me see my facial expression and mouth shape changes, and it caused more pressure''}.}

\paragraph{Pitch and volume with trends needed to be designed more carefully}
The effectiveness of pitch (M=4.00, SD=1.41), pitch trend (M=3.91, SD=1.44), volume (M=3.73, SD=1.54) and volume trend (M=3.91, SD=1.50) are not significant. Some participants found the pitch and volume values not representative or generalizable (\textit{P2: `I think pitch and volume depend on the situation you are speaking in''}). In the meantime, some participants preferred to see a qualitative indication, which should be the result of the comparison between their data and the optimal values (volume: between 50 and 65 decibels \cite{robbins_2022}; pitch: 100-120 Hz for adult male and higher for adult female \cite{microphones_2021}), similar to what they have suggested for speech rate (\textit{P3: ``Only show fast/slow/high/low instead of the number'' and P4: ``Pitch and volume are of little use, I need a reference value to know whether I am speaking at a high or low volume and pitch''}).

\sout{
We can clearly see from Figure~\ref{fig:daily-avg} that the average score for Pitch ($mean \ = \ 0.14$) \& Pitch Trend ($mean \ = \ 0.13$), Volume ($mean \ = \ 0.11$) \& Volume Trend ($mean \ = \ 0.10$) from daily questionnaires are in the bottom four of all real-time indicators and are far from both the average score and the standard deviation of the other types of feedback, which indicates that the effectiveness of these features are not significant, and the participants all tended to agree with this view (\textit{P2: `I think pitch and volume depend on the situation you are speaking in''}). In the meantime, some participants stated that they would prefer to see a qualitative indication, which should be the result of the comparison between their data and the optimal values (volume: between 50 and 65 decibels \cite{robbins_2022}; pitch: 100-120 Hz for adult male and higher for adult female \cite{microphones_2021}), similar to what they have suggested for speech rate (\textit{P3: ``Only show fast/slow/high/low instead of the number'' and P4: ``Pitch and volume are of little use, I need a reference value to know whether I am speaking at a high or low volume and pitch''}).
}

\sout{
\textbf{Motivating Methods}
Here we report the utilization of features designed to encourage users to persist in practicing daily, including the daily practice reminder, encouraging remarks on each page, ranking and daily report. The feedback towards these motivating features vary among different individuals, indicating that PWS have highly personalized training patterns. 

\paragraph{Reminder for Daily Practice}
As aforementioned in Section \ref{interfaceDesign}, a daily reminder function is provided to make sure participants engage in the training process in a timely and consistent manner.Before the start of the study, each participant indicated the time at which they would be reminded via WeChat to complete the daily training and comments. At the end of the study, participants were interviewed about three aspects: the effect of the reminders, the frequency of the reminders, and alternative potential reminder methods.

For the effect of the reminders, 7 out of 11 participants indicated that reminder was useful and necessary (\textit{P6: ``It stops my laziness''}). P5 and P7 reported that reminders were only essential for the first few days, and later they were able to remember on their own (\textit{P5: ``It was helpful at the beginning, but I started to remember to do it by myself later''}). In the meantime, 4 participants claimed that they do not need a reminder because their practice time fluctuates greatly with their daily free time (\textit{P8: ``It depends on personal consciousness and amount of available time''}). For the reminding frequency, four participants agreed that the frequency of once per day is sufficient (\textit{P1: ``The frequency is OK, but the reminder should be integrated to the system to let users set reminding frequency''}). Two participants would like to increase the frequency to twice a day (\textit{P5: ``It can be once in the morning and once before bed''}). P7 expressed a desire to reduce the frequency to once every three days (\textit{P7: ``Reminders are too frequent; once every three days would be sufficient''}). About other potential reminding ways, seven participants concurred that the reminding way employed in the study, a reminder via WeChat group or private chat messages, is effective (\textit{`` P8: WeChat messages are convenient. Phone call and application notifications are not recommended because they are disruptive’’}). P3 also recommended WeChat Public Account\footnote{WeChat Public Account: https://mp.weixin.qq.com/} (\textit{P3: ``Use a public account and broadcast the reminding messages''}). P4 and P5 believed that native app notifications would be preferable (\textit{P4: ``Notification of app is great, which I can click to enter today's tasks''}).
}

\section{Discussion}
\rv{
We summarize our key contributions and then discuss our design considerations and future work in light of prior work.  %novel findings compared to prior work, influence of cultural context and future works.

\subsection{Key Contributions}
We uncovered the stuttering scenarios, workarounds, and challenges of PwS in China by first analyzing four popular online forums where Chinese PwS actively participate and then conducting semi-structured interviews. Specifically, we identified the need for assistive mobile applications, general and specific Chinese stuttering situations, and the need for personalized timely feedback on their practice sessions and social support. From there we derived five design considerations: (1) The practice should focus on their stutter-triggering scenarios. (2) The practice scenarios should be carefully designed to trigger pressure and nervousness. (3) The practice scenarios should trigger users to practice stutter-triggering syllables, either implicitly or explicitly. (4) The platform should build a supportive community where PwS could give and receive timely personalized peer feedback on their practice sessions. (5) The platform should provide timely objective feedback on PwS's practice sessions for them to reflect on and adjust accordingly (e.g. speech rate and volume). 

We implemented these design considerations in a prototype named \sys{}  and evaluated it with a deployment study where 11 PwS participants used it online simultaneously for seven days. We derived key findings and insights to inform future design: (1) Stressful stutter-triggering practice scenarios close to daily lives could help PwS practice speaking fluently in real-life. (2) Timely subjective personalized peer feedback on their practice sessions could help PwS reflect on their speech by pointing out problems and providing advice, encouragement, and communication opportunities. (3) Real-time objective feedback could help PwS adjust their speech rate and facial expression. Next, we discuss each of our key findings in light of prior work and the design implications.

\subsection{Practice Scenarios}
Based on our design considerations, we summarized key findings about practice scenarios in three aspects: task topic, stutter-triggering syllables, and stress.

\paragraph{Task topic - \textbf{(DC1)}}
All participants agreed that practicing personalized tasks with stutter-triggering syllables in different stressful scenarios with our prototype could effectively help them improve their speech fluency and deal with similar situations in their daily lives. Although prior work uncovered potential task topics for PwS, only the usage of a subset of six topics was investigated  \cite{McNaney2018stammerApp}. Our work included all these six topics and added more Chinese-specific scenarios, such as reporting identity information when doing community-level PCR tests and reporting verification codes for receiving delivery. Moreover, we also observed personalized preferences toward different practice task topics and the number of conversation rounds in them. Thus, it is important to allow users to tailor their practice tasks based on preferences. Moreover, cultural context should also be considered as some scenarios (e.g. social events-related scenarios and living habits in an area) might only work in specific cultural contexts.

\paragraph{Stutter-triggering syllables - \textbf{(DC2)}}
Our formative study findings show that many Chinese PwS have persistent hard-to-pronounce and stutter-triggering syllables, many of which (e.g. `da' and `li') were not reported in prior work \cite{lea2021sep, McNaney2018stammerApp}. This might be caused by language differences and pronunciation habits. By embedding those stutter-triggering words in the conversation design of \sys{}, we found that participants stuttered on those words and were generally satisfied with the design. Furthermore, our formative study found that merely embedding stutter-triggering words in the conversation design was not enough because some participants thought practicing speaking a single sentence with their stutter-triggering words would be more efficient practice. Therefore, we designed some tasks without video which only asked the user to read the sentence on the screen aloud. However, our evaluation results indicated that none of the participants stuttered on these reading tasks. This does not mean that speaking a single sentence with stutter-triggering words was not beneficial for practice speaking, but only suggested that reading sentences on screen is not an effective way to train PwS to speak their stutter-triggering words. Future work should investigate better ways of designing stutter-triggering sentence practices.

\paragraph{Stress and realism - \textbf{(DC3)}}
Prior work found that PwS would like to practice speaking under pressure \cite{McNaney2018stammerApp}. However, it remained unknown how much stress PwS would prefer to have. Another novel design of \sys{}, As aforementioned in Sec \ref{finalPracticer}, was using the tasks to induce different stress levels by controlling the person in the video to make direct eye contact with the user or interrupt them when they were speaking, etc. According to our findings, different individuals had various preferences for this design. Some participants preferred tasks with high-stress levels while some preferred low-stress levels. Interestingly, some participants even found the highest stress level we offered was still not stressful enough. Furthermore, participants felt that the realism of the current task was limited by its format, a pre-recorded 2D video, and could not perfectly simulate the daily conversation with stress. To make the practice experience more realistic, future research should explore using other techniques, such as VR and AR, to provide a more immersive practice environment. In addition, face-swap techniques might be leveraged to make the practice condition more realistic or stressful by showing the face of someone whom they are afraid of speaking to in the practice task videos.

\subsection{Subjective Personalized Peer Feedback - \textbf{(DC4)}}
All PwS participants agreed that personalized timely subjective feedback on their practice sessions was beneficial for their training. Participants appreciated their peers' comments and advice on their problems, kind compliments, and encouragement (\textbf{DC4}). Although prior work also involved the monitor and feedback mechanism, most of them focused on registering the stutter situations in real life for SLP to better diagnose \cite{madeira2013building, demarin2015impact, madeira2018personalising} or practice independently without others' feedback \cite{McNaney2018stammerApp}. None of them investigated the effectiveness of peer feedback on their practice sessions, which is important to Chinese PwS due to the shortage of SLP in China \cite{ming2001public}. Prior work revealed that Chinese people have more negative attitudes towards their role in assisting PwS and empathy \cite{ustun2021cultural}, which may result in Chinese PwS feeling lonely and eager to build relationships with other PwS. This aligns with our findings on the lack of personalized peer support and advice in online communities. Therefore, we required participants to comment friendly while gently pointing out their issues. All participants felt the comments helped them and several participants reported receiving psychological encouragement. Although encouraging comments and compliments are heart-warming, some participants found some comments perfunctory, and they would like to get practical advice rather than perfunctory compliments. A prospective ideal comment should not only avoid reiterating a problem that has already been stated too many times for the sake of encouragement but also highlight important issues. Future work should analyze the characteristics of the comments in the PwS community and design learning-based AI models to moderate the comments in online communities and forums. The comments could be classified and labeled to make it more efficient for users to retrieve the information they want.

\subsection{Real-time Objective Feedback - \textbf{(DC5)}}
We also confirmed that many objective feedback features are effective (\textbf{DC5}), such as facial expression, speech rate, and transcription. Although prior work indicated speech-related indicators were beneficial for general speech training, how they would be utilized and perceived by PwS was not investigated \cite{fung2015roc}. We summarized the participants' usage patterns for each feature. Although participants found many types of objective feedback beneficial, many found it challenging to view multiple indicators simultaneously while practicing. Instead, they preferred to see qualitative labels (e.g. fast or slow) of their speech rather than the exact values while practicing. Thus, future work should investigate the most efficient presentation of various forms of objective information to assist real-time practice. While for reflection after practice, many participants preferred such detailed objective information that could help them reflect on their performance. These low rating scores of the current visualizations of features suggested that viewing trends of multiple features in parallel was overwhelming. Future work should investigate better ways to visualize such information to avoid information overloading. %Although we have offered the function to hide each feedback indicator, few of them were utilized in the study, which indicates that the visualization design should not be handled all by users. 
In addition, there are many other features that users expect, such as a summarization of common characteristics in users' stutter situations, straightforward indicators of where users were blocked, and a breathing indicator as some PwS tend to stutter after they speak for a long time and forget to inhale.
}

\section{Conclusion and Future Work}
In this work, we investigated the effectiveness of targeted practice scenarios and personalized timely feedback in assisting PwS in coping with stuttering and improving speech fluency. Informed by the literature and a formative study investigating user experiences and needs of PwS in China, we developed \sys{}, an online support tool for PwS to practice speech fluency and receive timely feedback, through an iterative design process involving twelve PwS. A seven-day deployment study with eleven PwS simultaneously online in China revealed that these key features could assist PwS in practicing to improve speech fluency, maintaining a positive mindset and facing similar situations in real life. In addition, we provided design implications for future work in assisting PwS to enhance their speech fluency.

%\subsection{Future Work}
To our knowledge, our work is the first to design and evaluate a mobile app that integrates a rich set of personalized practice scenarios, real-time speech indicators and timely peer feedback from real persons through a long-term multi-user simultaneous online study. %Future work could improve the design of some specific features with the aforementioned implications derived from our findings. 
Although the seven-day deployment study allowed us to derive insights into the effectiveness and user perception of the key features, the study time is still relative short to capture the whole picture of users' usage patterns. Future work should conduct an even longer deployment study with a revised mobile app that integrates the design implications from this work. 
%one participant expressed his like for this prototype and wish to be able to continue using it right after the study. 
%A longer term study could be done if future researchers would like to investigate the long term effect of these key features.

\begin{acks}
% This research was supported by xxx. 
We thank our reviewers for their constructive feedback and PwS participants for their participation.
\end{acks}

\bibliographystyle{ACM-Reference-Format}
\bibliography{main}

\appendix









\end{document}